\documentclass[aps,twocolumn,showpacs,superscriptaddress]{revtex4}

\usepackage{graphicx}
\usepackage{amssymb}

\usepackage{color}   
\usepackage{hyperref}  
\hypersetup{colorlinks,urlcolor=blue,linkcolor=blue,citecolor=blue,filecolor=blue,urlcolor=blue}

\usepackage[latin1]{inputenc}

\newcommand{\newc}{\newcommand}
\newc{\beq}{\begin{equation}}
\newc{\eeq}{\end{equation}}
\newc{\bea}{\begin{array}}
\newc{\eea}{\end{array}}
\newcommand{\ben}{\begin{eqnarray}}
\newcommand{\een}{\end{eqnarray}}
\newc{\ra}{\rightarrow}
\newc{\bfx}{{\bf x}}
\newc{\bfV}{{\bf V}}
\newc{\cO}{{\cal O}}
\newc{\bfv}{{\bf v}}
\newc{\bfu}{{\bf u}}
\newc{\bfp}{{\bf p}}
\newc{\ve}{{\varepsilon}}
\newc{\Psibar}{\overline\Psi}
\newc{\w}{{\bf w}}
\newc{\E}{{\mathbf{E}}}
\newc{\EE}{{\mathcal E}}
\newc{\bfn}{{\mathbf\nabla}}
\newc{\la}{{\cal L}}
\newc{\tla}{{\tilde{\cal L}}}
\newc{\bp}{{\bf p}}
\newc{\ho}{\hookrightarrow }
\newc{\bP}{{\bf P}}
\newc{\pd}{{\partial}}
\newc{\piv}{{\partial_4}}
\newc{\pv}{{\partial_5}}
\newc{\bJ}{{\bf J}}
\newc{\bze}{{\mathbf 0}}
\newc{\bK}{{\bf K}}
\newc{\tphi}{{\tilde\phi}}
\newc{\tF}{{\tilde F}}
\newc{\tD}{{\tilde D}}
\newc{\tJ}{{\tilde J}}
\newc{\tj}{{\tilde j}}
\newc{\bD}{{\bf D}}
\newc{\tvphi}{{\tilde\varphi}}
\newc{\trho}{{\tilde\rho}}
\newc{\ttheta}{{\tilde\theta}}
\newc{\tpsi}{{\tilde\psi}}
\newc{\tu}{{\tilde u}}
\newc{\cD}{{\cal D}}
\newc{\tPhi}{{\tilde\Phi}}
\newc{\tPsi}{{\tilde\Psi}}
\newc{\tA}{{\tilde A}}
\newc{\talpha}{{\tilde\alpha}}
\newc{\tbeta}{{\tilde\beta}}
\newc{\bA}{{\mathbf A}}
\newc{\bB}{{\bf B}}
\newc{\br}{{\bf r}}
\newc{\sig}{{\mathbf\sigma}}
\newc{\eg}{{\rm e.g.\ }}
\newc{\ie}{{\rm i.e.\ }}
\newcommand{\bey}{\begin{eqnarray}}
\newcommand{\pslash}{\not{\hbox{\kern-2.3pt $p$}}}
\newcommand{\pdslash}{\not{\hbox{\kern-2pt $\partial$}}}
\newcommand{\eey}{\end{eqnarray}}

\begin{document}

\title{ Majorana fermions, supersymmetry and thermofield dynamics}

\author{ Marco A. S. Trindade \footnote{\href{matrindade@uneb.br}{matrindade@uneb.br}} }

\affiliation{Departamento de Ci\^encias Exatas e da Terra, Universidade do Estado da Bahia, Colegiado de F\'isica, Bahia, Brazil.}

\author{Sergio Floquet}
\affiliation{Colegiado de Engenharia Civil, Universidade Federal do Vale do S\~ao Francisco, Juazeiro-BA, Brazil }

\date{\today}

\begin{abstract}
In this work we show the existence of supersymmetry and degeneracy for an arbitrary number of Majorana fermions (even or odd) without to invoke any symmetry of Hamiltonian. Next, we analyze the supersymmetry at finite temperature using the thermofield dynamics formalism. Furthermore we derive thermal braiding operators through the Bogoliubov transformations and we find its action on a thermal Bell state.

\vspace{0.5cm}


\pacs{11.30 Pb; 03.65. Fd; 11.10.wx; 74.25-q}

\end{abstract}

\maketitle




\section{Introduction}
Majorana fermions are quasiparticles that might to emerge in superconducting materials as delineated in the seminal work of Kitaev \cite{Kitaev1}. The Kitaev's model generate a doubled spectrum for the Majorana zero modes. These degenerated states has applications for a proposal of topological quantum computer due to non-abelian statistics of the bound states \cite{Alicea}. The topological quantum computation is based on the quasiparticles know as non-Abelian anyons \cite{Yu,Long}, whose excitations satisfy non-Abelian braid statistics \cite{Yu,Long} and it produces a computation scheme immune to errors \cite{Kitaev2, Kitaev3}. The existence of degenerate ground states plays a pivotal role since under local pertubations the system evolues only within the ground state subspace characterizing an fault-tolerant scheme \cite{ Alicea, Kitaev2, Kitaev3, Nayak}.   \

Lee and Wilczek \cite{Lee} analyze the subjacent algebraic structure of the doubled spectrum. It has been shown that for a junction supporting an odd number of Majorana mode operators, there is an emergent operator that leading to degeneracy. In addition, the doubling applies to all energy eigenstates, not restricted to ground state. An interesting discussion about underlying emergent supersymmetry is performed  in the reference \cite{Qi} where there is an emergent supersymmetry arise from time-reversal symmetry. Subsequently, Hsieh \emph{et al}. \cite{Hsieh} showed that for Majorana models with translation symmetry, the supersymmetry leads to spectrum doubling. The supersymmetry associated to the doubling Majorana algebra \cite{Haq} is explicitly derived. \

One of the seminal works on supersymmetry at finite temperature was carried out by Van Hove \cite{Hove}. He studied how supersymmetry is related to the properties of excited states and derive relations that extend the typical consequences of supersymmetry for ground-state expectation values. Supersymmetry in thermofield dynamics has been studied in \cite{Par}. Mathematical possibilities of preserving supersymmetry were presented and the spontaneous breakdown of supersymmetry were investigated in comparison with Van Hove's previous work. \cite{Hove}

In this work we present an underlying supersymmetry algebra for Majorana modes that applies to an even or odd number of Majorana modes operators. Then we show that this supersymmetry can be extend to a finite temperature formulation using thermofield dynamics prescription \cite{Khanna}. The paper is structured as follows. In Section \ref{sec2} we discussed the relationship between supersymmetry and degeneracy. In Section \ref{sec3} we present a formulation for Majorana fermions using the TFD formalism. Section \ref{sec4} contains a generalization to an arbitrary number of Majorana modes and the construction of thermal braiding operators. In Section \ref{sec5} we have the conclusions and perspectives.

\section{Supersymmetry and degeneracy \label{sec2}}
Consider a chain with $N$ ordered sites whereupon each site can be either empty or occupied by an electron. The Kitaev's fire model is defined by Hamiltonian \cite{Kitaev1}
\begin{eqnarray}
H' & = & \sum_{j=1}^{N}  \left[-\phi(a_{j}^{\dag}a_{j+1}+a_{j+1}^{\dag}a_{j})-\mu(a_{j}^{\dag}a_{j}-\frac{1}{2}) \right. \nonumber \\
& & \left. + \Delta a_{j}a_{j+1} + \Delta^{\ast}a_{j+1}^{\dag}a_{j}^{\dag} \right],
\end{eqnarray}
where $\phi$ is a hopping amplitude, $\mu$ is a chemical potential and $\Delta$ the superconducting gap. We can rewrite this Hamiltonian defining the Majorana operators
\begin{eqnarray}
\gamma_{2j-1}=a_{j}+a_{j}^{\dag}; \ \ \gamma_{2j}=\frac{a_{j}-a_{j}^{\dag}}{i},
\end{eqnarray}
so that we have Clifford algebra
\begin{equation}
\{\gamma_{k},\gamma_{l}\}=2\delta_{k,l},
\end{equation}
and the rewritten Hamiltonian is
\begin{eqnarray}
H' & = & \frac{i}{2}\sum_{j=1}^{N} \left[-\mu \gamma_{2j-1}\gamma_{2j}+(\phi+|\Delta|)\gamma_{2j}\gamma_{2j+1} \right. \nonumber \\
& & \left. +(-\phi+|\Delta|)\gamma_{2j-1}\gamma_{2j+2} \right] . \label{eq-4}
\end{eqnarray}
 The particular of equation (\ref{eq-4}) case $\mu=0$; $|\Delta|=\phi >0$  is given by
\begin{eqnarray}
H' & = & i \phi \sum_{j=1}^N \gamma_{2j}\gamma_{2j+1},
\end{eqnarray}
and note that operators $\gamma_{1}$ e $\gamma_{2N}$ do not appear in $H'$. Kitaev \cite{Kitaev1} shown that ground state is degenerate.

We consider the effective Hamiltonian
\begin{eqnarray}
H & = & -i \phi \gamma_{1}\gamma_{2N},
\end{eqnarray}
with the interactions on the two ends of wire. We shown that there is a supersymmetry algebra related to hamiltonian, which implies degeneracy of system. A supersymmetry algebra is a Lie superalgebra. One can define Lie superalgebra over $\mathbb{R}$ or $\mathbb{C}$ as a $Z_{2}$-graded algebra $g=g_{0}\bigoplus g_{1}$, whose supercommutator satisfies the following conditions \cite{Kac,Traubenberg}

\begin{itemize}
 \item[1)] Super skew-symmetry
\begin{eqnarray}
[X,Y]=-(-1)^{|X||Y|}[Y,X]
\end{eqnarray}

\item[2)] Super Jacobi identity
\begin{eqnarray}
(-1)^{|X||Y|}[X,[Y,Z]]+(-1)^{|Z||Y|}[Y,[Z,X]]\nonumber \\
+(-1)^{|Z||Y|}[Z,[X,Y]]=0
\end{eqnarray}
where $|X|$ is the degree of $X$. 

\end{itemize}

 We should have \cite{Traubenberg} $[g_{0},g_{0}]\subseteq g_{0}$, $[g_{0},g_{1}]\subseteq g_{1}$ and  $\{g_{1},g_{1}\} \subseteq g_{0}$ ($\{ \ , \ \} $  is the symmetric product). Consider the basis of $g_{0}$ and $g_{1}$, $B_ {g_{0}}=\{1,H\}$ and $B_{g_{1}}=\{Q_{1}=\gamma_{2},Q_{2}=\gamma_{2L-1}\}$ associated to the effective Hamiltonian with $L\equiv N$. We have
\begin{eqnarray}
 \hspace{-1cm}\left\lbrace \begin{array}{l}
[Q_{i},H] \ = \ 0,  \\
\{Q_{i},Q_{j}\} \  = \ \{Q_{i},Q_{j}^{\dag}\}  \ = \ \{Q_{i}^{\dag},Q^{\dag}_{j}\} \ = \ 2\delta_{ij}1 ,
\end{array} \right.  \ i, j=1,2 \nonumber \hspace{-1cm} \\
\end{eqnarray}
wherein $Q_{1}^{\dagger} = Q_{1}$ and $Q_{2}^{\dagger} = Q_{2}$. Interestingly, we note that
\begin{eqnarray}
[P,H]=\{P,Q_{i}\}=0,
\end{eqnarray}
where $P=\gamma_{1}\gamma_{2}\gamma_{2L-1}\gamma_{2L}$ is the parity operator. We can now to prove the degeneracy. Consider the ground state $|\psi_{0}\rangle$ and note that $Q_{i}|\psi_{0}\rangle$ and $|\psi_{0}\rangle$ are orthogonals. In fact,
\begin{eqnarray}
\langle \psi_{0}|Q_{i}|\psi_{0}\rangle & = & \langle \psi_{0}|Q_{i} \left( \pm P \right) |\psi_{0}\rangle \nonumber \\
& = & - \langle \psi_{0}|\left( \pm P \right) Q_{i}|\psi_{0}\rangle \nonumber \\
& = & -\langle \psi_{0}|Q_{i}|\psi_{0}\rangle \nonumber \\
& = & 0.
\end{eqnarray}

It is important to highlight that $Q_{i}|\psi_{0}\rangle \neq 0$ ($Q_{i}^{2}= 1$). The supercharges $Q_{i}$ change the parity of state. If $|\psi_{0}\rangle$ has even parity, $Q_{i}|\psi_{0}\rangle$ has odd parity.

\section{TFD for the Majorana fermions \label{sec3}}

For an analysis at finite temperature a thermofield dynamics approach can be constructed from the notion of w*-algebras and Tomita-Takesaki theory \cite{Tak, Ojima, Bratteli, Santana}. Let $\mathcal{H}$ be a Hilbert space and $ \pi(A)$ a faithful representation a c*-algebra $A$. Considering $ \vert \varphi \rangle \in \mathcal{H}$, we have that $\langle \varphi  \vert \pi (A) \vert \varphi \rangle$ defines a state over $A$ given by $ \omega_{\varphi}(A)=\langle \varphi \vert \pi (A) \vert \varphi \rangle$. Inversely, by the GNS construction \cite{Santana}, every state $\omega$ of $A$ has a vector $\vert \varphi_{\omega} \rangle \in \mathcal{H}$ such that $\omega(A)=\langle \varphi \vert \pi_{\omega} (A) \vert \varphi \rangle$ .

A w*-algebra is a c*-algebra that is weakly closed and contains the identity. Let $J:\mathcal{H}\rightarrow \mathcal{H}$ an antilinear isometric involution with $J^{2}=1$. We have a Tomita-Takesaki representation of the w*-algebra iff $J \pi_{\omega}(A)J=\widetilde{\pi_{\omega}}(A)$ define an *-antiisomorphism on the linear operators, where $\widetilde{\pi_{\omega}}(A)$ is the commutant of $\pi_{\omega}(A)$, i.e., $[\pi_{\omega}(A),\widetilde{\pi_{\omega}}(A)]=0$. We will denote the elements of $\pi_{\omega}(A)$ by $A$ and the elements of $\widetilde{\pi_{\omega}}(A)$ by $\widetilde{A}$ \cite{Santana}. For the fermionic annihilation and creation operators we have $\pi_{\omega}(a)=a$, $\widetilde{\pi_{\omega}}(a)=\widetilde{a}$, $\pi_{\omega}(a^{\dag})=a^{\dag}$, $\widetilde{\pi_{\omega}}(a^{\dag})=\widetilde{a^{\dag}}$, $\widetilde{a}=JaJ$ and $\widetilde{a^{\dag}}=Ja^{\dag}J$.

Following the TFD prescription \cite{Khanna} we introduced the thermal Majorana operators $\gamma_{2j-1}(\beta)$ and $\gamma_{2j}(\beta)$ through the relations
\begin{eqnarray}
\gamma_{2j-1}(\beta)&=&U(\beta)(a_{j}+a_{j}^{\dag})U^{\dag}(\beta) \nonumber \\
&=&U(\beta)a_{j}U^{\dag}(\beta)+U(\beta)a_{j}^{\dag}U^{\dag}(\beta) \nonumber \\
&=&a_{j}(\beta)+a_{j}^{\dag}(\beta),
\end{eqnarray}
\begin{eqnarray}
\gamma_{2j}(\beta)&=&U(\beta)i(a_{j}^{\dag}-a_{j})U^{\dag}(\beta) \nonumber \\
&=&iU(\beta)a_{j}^{\dag}U^{\dag}(\beta)-U(\beta)a_{j}U^{\dag}(\beta) \nonumber \\
&=&i[a_{j}^{\dag}(\beta)-a_{j}(\beta)],
\end{eqnarray}
\begin{eqnarray}
\widetilde{\gamma_{2j-1}}(\beta)&=&U(\beta)(\widetilde{a_{j}}+\widetilde{a_{j}}^{\dag})U^{\dag}(\beta) \nonumber \\
&=&U(\beta)\widetilde{a_{j}}U^{\dag}(\beta)+U(\beta)\widetilde{a_{j}}^{\dag}U^{\dag}(\beta) \nonumber \\
&=&\widetilde{a_{j}}(\beta)+\widetilde{a_{j}}^{\dag}(\beta),
\end{eqnarray}
and
\begin{eqnarray}
\widetilde{\gamma_{2j}}(\beta)&=&U(\beta)i(\widetilde{a_{j}}^{\dag}-\widetilde{a_{j}} )U^{\dag}(\beta) \nonumber \\
&=&iU(\beta)\widetilde{a_{j}}^{\dag}U^{\dag}(\beta)-U(\beta)\widetilde{a_{j}}U^{\dag}(\beta) \nonumber \\
&=&i[\widetilde{a_{j}}^{\dag}(\beta)-\widetilde{a_{j}}(\beta)],
\end{eqnarray}
where $U(\beta)=\exp[\theta(a^{\dag}\widetilde{a}^{\dag} - \widetilde{a}a)]$, $\displaystyle \cos(\theta)=\frac{1}{\sqrt{1+e^{-\beta \phi}}}$ and $ \sin(\theta) = \displaystyle \frac{1}{\sqrt{1+e^{\beta \phi}}}$, with commutation relations
$\{\widetilde{a}^{\dag},\widetilde{a}\}=1$ and $\{\widetilde{a}^{\dag},\widetilde{a}^{\dag} \}=\{\widetilde{a},\widetilde{a}\}=\{a,\widetilde{a}\}=\{\widetilde{a}^{\dag},a\}=\{a^{\dag},\widetilde{a}\}=
\{a^{\dag},\widetilde{a}^{\dag}\}=0$ and $\displaystyle \beta= \frac{1}{ K_{b}T}$,  where $T$ is the temperature of the system in thermal equilibrium and $K_{b}$ is the Boltzmann constant.

 Apply the TFD in our construction we have
\begin{eqnarray}
\left[ Q_{i}(\beta),H(\beta) \right] & = & \left[ \widetilde{Q_{i}}(\beta),\widetilde{H}(\beta) \right] \ = \ \left[  Q_{i}(\beta),\widetilde{H}(\beta) \right]  \nonumber \\
& = & \{Q^{\dag}_{i}(\beta),\widetilde{Q}^{\dag}_{j}(\beta)\} \ = \ \{Q_{i}(\beta),\widetilde{Q}^{\dag}_{j}(\beta)\} \nonumber \\
& = & \{Q_{i}(\beta),\widetilde{Q_{j}}(\beta)\} \ = \ 0 , \nonumber \\
\{Q_{i}(\beta),Q_{j}(\beta)\} & = & \{Q^{\dag}_{i}(\beta),Q^{\dag}_{j}(\beta)\}= \{Q_{i}(\beta),Q^{\dag}_{j}(\beta)\}    \nonumber  \\
& = & \{\widetilde{Q}^{\dag}_{i}(\beta),\widetilde{Q}^{\dag}_{j}(\beta)\}= \{\widetilde{Q_{i}}(\beta),\widetilde{Q}^{\dag}_{j}(\beta)\} \nonumber \\
& = & \{\widetilde{Q_{i}}(\beta),\widetilde{Q_{j}}(\beta)\} = 2\delta_{ij}1,
\end{eqnarray}
and
\begin{eqnarray}
\left[ P(\beta),H(\beta) \right] & = & \{P(\beta),Q_{i}(\beta)\}=0, \nonumber \\
\left[ \widetilde{P}(\beta),\widetilde{H}(\beta) \right] & = & \{\widetilde{P}(\beta),\widetilde{Q_{i}}(\beta)\}=0, \nonumber \\
\left[ P(\beta),\widetilde{H}(\beta) \right] & = & \{P(\beta),\widetilde{Q_{i}}(\beta)\}=0,
\end{eqnarray}
with
\begin{eqnarray}\left\lbrace \begin{array}{rcl}
Q_{i}(\beta)&=&U_{k}(\beta)Q_{i}U_{k}^{\dag}(\beta)  \\
Q^{\dag}_{i}(\beta)&=&U_{k}(\beta)Q^{\dag}_{i}U_{k}^{\dag}(\beta)  \\
H(\beta)&=&U_{1}(\beta)U_{L}(\beta)HU_{1}^{\dag}(\beta)U_{L}^{\dag}(\beta)  \\
P(\beta)&=&U_{1}(\beta)U_{L}(\beta)PU_{1}^{\dag}(\beta)U_{L}^{\dag}(\beta)  \\
\end{array} \right. \end{eqnarray}
and
\begin{eqnarray}
\left\lbrace \begin{array}{rcl}
\widetilde{Q_{i}}(\beta)&=&U_{k}(\beta)\widetilde{Q_{i}}U_{k}^{\dag}(\beta)  \\
\widetilde{Q^{\dag}_{i}}(\beta)&=&U_{k}(\beta)\widetilde{Q}^{\dag}_{i}U_{k}^{\dag}(\beta)  \\
\widetilde{H}(\beta)&=&U_{1}(\beta)U_{L}(\beta)\widetilde{H}U_{1}^{\dag}(\beta)U_{L}^{\dag}(\beta)  \\
\widetilde{P}(\beta)&=&U_{1}(\beta)U_{L}(\beta)\widetilde{P}U_{1}^{\dag}(\beta)U_{L}^{\dag}(\beta)  \\
\end{array} \right.
\end{eqnarray}
where $k=\{ 1,L \}$. Therefore $|\psi_{0}({\beta})\rangle\equiv U_{k}(\beta)|\psi_{0}, \widetilde{\psi_{0}}\rangle$ and $Q_{i}(\beta)|\psi_{0}({\beta})\rangle$ are orthogonals, analogously to non-thermal case:
\begin{eqnarray}
\langle \psi_{0}({\beta})|Q_{i}({\beta})|\psi_{0}({\beta})\rangle &=&
\langle \psi_{0}({\beta})|Q_{i}({\beta}) \left( \pm P({\beta}) \right) |\psi_{0}({\beta})\rangle \nonumber \\
&=& - \langle \psi_{0}({\beta})| \left( \pm P({\beta}) \right)Q_{i}({\beta})|\psi_{0}({\beta})\rangle \nonumber \\
&=& -\langle \psi_{0}({\beta})|Q_{i}({\beta})|\psi_{0}({\beta})\rangle \nonumber \\
&=& 0.
\end{eqnarray}

Note that at limit $T\rightarrow 0$ we will get the results from the previous section.

\section{Thermal Generalization for an arbitrary number of Majorana modes and Braiding operators \label{sec4}}

The preceding development is valid for any state and not only for the ground state. We can be generalized these results for a Hamiltonian with an arbitrary number of thermal Majorana operators:
\begin{eqnarray}
H(\beta) &=&-i\alpha_{1,2L_{1}}\gamma_{1}(\beta)\gamma_{2L_{1}}(\beta)-i\alpha_{2L_{1},2L_{2}}\gamma_{2L_{1}}(\beta)\gamma_{2L_{2}}(\beta) \nonumber \\
& & - ... -i\alpha_{2L_{N}-1,2L_{N}}\gamma_{2L_{N}-1}(\beta)\gamma_{2L_{N}}(\beta) \nonumber \\
&=&-i\alpha_{1,2L_{1}}\gamma_{1}(\beta)\gamma_{2L_{1}}(\beta) \nonumber \\
 & & -i\sum_{k=1}^{N-1}\alpha_{2L_{k},2L_{k+1}}\gamma_{2L_{k}}(\beta) \gamma_{2L_{k+1}}(\beta) \nonumber \\
&=&\sum_{k=0}^{N-1}H_{k}(\beta),
\end{eqnarray}
where $H_{0}(\beta)=-i\alpha_{1,2L_{1}}\gamma_{1}(\beta)\gamma_{2L1}(\beta)$  and $H_{k}=-i\alpha_{2L_{k},2L_{k+1}}\gamma_{2L_{k}}(\beta)\gamma_{2L_{k+1}}(\beta)$.
Consider the basis of $g_{0}$ and $g_{1}$,  $B_{g_{0}}=\{1,H_{0}(\beta),H_{1}(\beta),...\}$ and $B_{g_{1}}=\{Q_{1,1}(\beta)\equiv \gamma_{2}(\beta) ,Q_{1,2}(\beta)\equiv \gamma_{2L_{1}-1}(\beta),Q_{2,1}(\beta)\equiv \gamma_{2L_{1}+1}(\beta),Q_{2,2}(\beta)\equiv \gamma_{2L_{2}-1}(\beta),...,Q_{N,1}(\beta)\equiv \gamma_{2L_{N-1}+1}(\beta), Q_{N,2}(\beta)\equiv \gamma_{2L_{N}-1}(\beta) \} $. The commutation relations are given by
\begin{eqnarray}
\left[ H_{a}(\beta),H_{b}(\beta) \right] & = & C_{ab}^{c}H_{c}(\beta), \nonumber  \\
\left[ H_{a}(\beta), Q_{b}(\beta) \right] & = & 0,
\end{eqnarray}
\begin{eqnarray}
\{Q_{a}(\beta),Q_{b}(\beta) \} & = &\{Q_{a}^{\dag}(\beta) ,Q_{b}^{\dag}(\beta) \}  \nonumber \\
& = & \{Q_{a}(\beta),Q_{b}^{\dag}(\beta)\}  \nonumber \\
& = & 2 \delta_{a,b},
\end{eqnarray}
and
\begin{eqnarray}
[H_{a}(\beta) ,P(\beta)]=\left\lbrace Q_{a}(\beta) ,P(\beta) \right\rbrace =0,
\end{eqnarray}
where
\begin{eqnarray}
P(\beta) & = & \gamma_{1}(\beta)\gamma_{2}(\beta)\gamma_{2L_{1}-1}(\beta)\gamma_{2L_{1}+1}(\beta)\gamma_{2L_{2}-1}(\beta) \nonumber \\
& & \times \gamma_{2L_{2}+1}(\beta) \gamma_{2L_{3}-1}(\beta)\gamma_{2L_{3}+1}(\beta)...\gamma_{2L_{N}-1}(\beta) \nonumber \\
& & \times \gamma_{2L_{N}+1}(\beta)
\end{eqnarray}
for N even and
\begin{eqnarray}
P & = & -i\gamma_{1}(\beta)\gamma_{2}(\beta)\gamma_{2L_{1}-1}(\beta)\gamma_{2L_{1}+1}(\beta)\gamma_{2L_{2}-1}(\beta) \nonumber \\
& & \times \gamma_{2L_{2}+1}(\beta)\gamma_{2L_{3}-1}(\beta) \nonumber \\
& & \times \gamma_{2L_{3}+1}(\beta)...\gamma_{2L_{N}-1}(\beta)\gamma_{2L_{N}+1}(\beta),
\end{eqnarray}
for N odd.

We can construct thermal braiding operators $R_{i}(\theta; \beta ) $ through the $R_{i}(\theta)=\exp(\theta \gamma_{i}\gamma_{i+1})$ \cite{Yu} and the Bogoliubov transformation $U(\beta)$. We notice that
\begin{eqnarray}
U(\theta)R_{i}(\theta) \vert \psi, \widetilde{\phi} \rangle &=& U(\beta)R_{i}(\theta) U^{\dagger}(\beta)U(\beta)\vert \psi, \widetilde{\phi} \rangle \nonumber \\
&=&R_{i}(\theta, \beta) \vert \psi, \widetilde{\phi}; \beta \rangle
\end{eqnarray}
and
\begin{eqnarray}
U(\theta)\widetilde{R_{i}}(\theta) \vert \psi, \widetilde{\phi} \rangle &=& U(\beta)\widetilde{R_{i}}\theta U^{\dagger}(\beta)U(\beta)\vert \psi, \widetilde{\phi} \rangle \nonumber \\
&=&\widetilde{R_{i}}(\theta, \beta) \vert \psi, \widetilde{\phi}; \beta \rangle ,
\end{eqnarray}
where
\begin{eqnarray}
R_{i}(\theta; \beta)&=&U(\beta)R_{i}(\theta)U^{\dagger}(\beta) \nonumber \\
\widetilde{R_{i}}(\theta; \beta)&=&U(\beta)\widetilde{R_{i}}(\theta)U^{\dagger}(\beta) ,
\end{eqnarray}
satisfying the relations
\begin{eqnarray}
\hspace{-0.3cm} R_{i}(\theta; \beta)R_{i+1}(\theta; \beta)R_{i}(\theta; \beta)&=&R_{i+1}(\theta; \beta)R_{i}(\theta; \beta)R_{i+1}(\theta; \beta) \nonumber \\
\hspace{-0.3cm}\widetilde{R_{i}}(\theta; \beta)\widetilde{R_{i+1}}(\theta; \beta)\widetilde{R_{i}}(\theta; \beta)&=&\widetilde{R_{i+1}}(\theta; \beta)\widetilde{R_{i}}(\theta; \beta)\widetilde{R_{i+1}}(\theta; \beta) . \nonumber \\ \hspace{-0.3cm}
\end{eqnarray}

Now we investigate the action of  the thermal braiding operator $R_{i}(\theta; \beta)$ on the $\vert \psi, \widetilde{\phi} \rangle$. Let us consider the Clifford algebra with thermal generators $\gamma_{1}(\beta)$,  $\gamma_{2}(\beta)$ and $\gamma_{3}(\beta)$, with thermal braiding operators
\begin{eqnarray}
R_{1}(\theta; \beta) & = & U(\beta)\exp(\theta \gamma_{1}\gamma_{2})U^{\dagger}(\beta) \nonumber \\
R_{2}(\theta; \beta) & = & U(\beta)\exp(\theta \gamma_{2}\gamma_{3})U^{\dagger}(\beta) .
\end{eqnarray}
 The action of thermal braiding operators on  a thermal Bell state given by
$\vert \psi(\beta) \rangle = a \vert 00\widetilde{00}(\beta)\rangle +b\vert 11\widetilde{00}(\beta)\rangle$, is
\begin{eqnarray}
R_{1}(\theta; \beta)\vert \psi(\beta) \rangle & = & \cos(\theta) \left[ a \vert 00\widetilde{00}(\beta)\rangle +b\vert 11\widetilde{00}(\beta)\rangle) \right. \nonumber \\ & &
\left. +\sin(\theta)a \vert 00\widetilde{00}(\beta)\rangle -b\vert 11\widetilde{00}(\beta)\rangle \right] \nonumber \\
\end{eqnarray}
and
\begin{eqnarray}
R_{2}(\theta; \beta)\vert \psi(\beta) \rangle & = &  \cos(\theta) \left[ a \vert 00\widetilde{00}(\beta)\rangle +b\vert 11\widetilde{00}(\beta)\rangle) \right. \nonumber \\ & &
\left. +\sin(\theta)a \vert 11\widetilde{00}(\beta)\rangle -b\vert 00\widetilde{00}(\beta)\rangle \right]. \nonumber \\
\end{eqnarray}

Notice that at $T\rightarrow 0$, the $\vert \psi(\beta) \rangle$ correspond to Bell state ($\vert \psi \rangle = a \vert 00 \rangle + b \vert 11 \rangle$  with $a=b=1/\sqrt{2}$) \cite{Khanna} and the action of $R_{1}(\theta; \beta)$ and $R_{2}(\theta; \beta)$ on this state preserves the entanglement.

\section{Conclusions \label{sec5}}
In this paper we present a supersymmetry algebra for Majorana modes that leading the degeneracy of spectrum. It is important to note that we have not invoked any additional symmetry for the system and that our results are valid for any number (even or odd) of Majorana mode operators. Besides our formulation supports an extension to finite temperature. We introduced thermal Majorana operators through Bogoliubov transformations. Then, we derive the thermal braiding operators and and we found that they preserve entanglement considering their actions on the thermal Bell states. As perspective we intend to calculate the thermal Green functions associated with these systems.  Another interesting aspect to be investigated is the breakdown of supersymmetry in this context.


\end{document}